# Mapping Philanthropic Support of Science

Louis M. Shekhtman[1], Alexander J. Gates[1], Albert-László Barabási[1,2,3,4]

[1]Network Science Institute, Northeastern University, Boston, Massachusetts 02115, USA

[2]Center for Cancer Systems Biology, Dana-Farber Cancer Institute, Boston, Massachusetts 02115, USA

[3]Department of Medicine, Brigham and Women's Hospital, Harvard Medical School, Boston, Massachusetts 02115, USA

[4]Department of Network and Data Science, Central European University, Budapest 1051, Hungary

**Abstract**

While philanthropic support for science has increased in the past decade, there is limited quantitative knowledge about the patterns that characterize it and the mechanisms that drive its distribution. Here, we map philanthropic funding to universities and research institutions based on IRS tax forms from 685,397 non-profit organizations. We identify nearly one million grants supporting institutions involved in science and higher education, finding that in volume and scope, philanthropic funding has grown to become comparable to federal research funding. Yet, distinct from government support, philanthropic funders tend to focus locally, indicating that criteria beyond research excellence play an important role in funding decisions. We also show evidence of persistence, i.e., once a grant-giving relationship begins, it tends to continue in time. Finally, we leverage the bipartite network of supporters and recipients to help us demonstrate the predictive power of the underlying network in foreseeing future funder-recipient relationships. The developed toolset could offer funding recommendations to organizations and help funders diversify their portfolio. We discuss the policy implications of our findings for philanthropic funders, individual researchers, and quantitative understanding of philanthropy.

**INTRODUCTION**

Since the emergence of a US federal funding system for research following World War II,[1] public sources of funding have failed to keep up with the growing demands of fundamental and applied research[2]. Philanthropy increasingly fills this gap, in 2016 contributing up to 44% of basic research funding at US universities[3,4], and is credited for high-impact outcomes such as supporting the work of Chemistry Nobel Prize recipients Frances Arnold and Jennifer Doudna[5]. While the patterns characterizing US federal funding of science and university research are closely monitored and are the subject of spirited policy debates[6-10], our understanding of the philanthropic ecosystem is often limited to summary statistics, case studies, or the largest gifts[3,11-16]. This narrow focus prohibits a quantitative understanding of the complete spectrum of philanthropic support for scientific institutions and is thus unable to identify systematic patterns that arise. Philanthropists and the community have become increasingly aware of these obstacles and have begun calling for increased research, both into science philanthropy and into philanthropic funding more generally[17-19]. The obstacles towards a quantitative understanding of philanthropy have primarily been rooted in data availability: while all details pertaining to federal funding are public and accessible for research purposes, we lack a similar transparency when it comes to philanthropic grant-giving.

Data access has fortunately improved recently thanks to changes by the Internal Revenue Service (IRS), who has made machine-readable Form 990 tax data available for research[20]. This tax form is filed by all US non-profits and foundations (except churches), and contains information on the organization's revenue, expenditures, executive leadership[21], mission statement[22,23], and more. We used this resource to analyze over 3.6 million tax forms from 685,397 non-profit organizations in the United States between 2010-2019, extracting information about over 10 million grants (Fig. S1). The data allowed us to identify 69,675 nonprofit organizations involved in funding or performing scientific research, who together gave and received 926,124 grants totaling $208 billion. We find that funding offered by philanthropy to research institutions has reached $30B per year in recent years, rivaling the level of funding offered by the NIH (Fig. 2a).

Our analysis reveals important differences between philanthropic and federal funding. While the government relies on a few large organizations to fund scientific research, as we show, the philanthropic ecosystem is extremely heterogenous and distributed, where a few large foundations coexist with many small funders. Similarly, we find that philanthropic grants have distinct patterns in terms of both geography and temporality. Philanthropic funders have a strong preference for funding local organizations and, in contrast to federal grants that tend to last for defined terms, philanthropic relationships become increasingly entrenched over time. Finally, we leverage the network patterns of funders and recipients to predict which funders are most likely to support a given recipient, empowering us to offer funding recommendations to organizations and help funders diversify their portfolio.

**PHILANTHROPIC FUNDING OF SCIENCE**

US-based nonprofit organizations are required to file Form 990, detailing their executive leadership, assets, cash flow, and layers of financial information, collected and publicly shared by the IRS. Nonprofit organizations self-report their area of activity under 26 categories, four of which have direct relevance to science: Social Science Research Institutes; Science and Technology Research Institutes; Medical Research; and Higher Education Institutions, together representing 30,351 organizations. We added to this list 3,738 public universities and nonprofit organizations that receive grants on their behalf (see SI Section I). After identifying all grants received by these organizations, we arrive to 69,675 non-profit organizations who have donated or received funds that contribute to science. In Fig. 1, we show the network of 1,254 funders who gave at least $1M to one of 55 recipients. The nodes are colored by US region, unveiling the strong preference of funders for recipients in their same region.

We find a clear distinction between funders and recipients: 81% of the identified organizations only gave grants, 16% only received grants, and 3%, mainly universities, were involved in both giving and receiving. The fact that the number of funders significantly exceeds the number of recipients is somewhat unexpected, given the difficulty many organizations report in attracting funding[24]. However, it is worth noting that each university and many research institutions contain multiple departments and research groups that independently seek funding. Thus, rather than evidence of a plethora of funding opportunities available for scientists, the imbalance between donor and recipient organizations reflects the fact that most research is carried out within a few large institutions, mainly universities, that provide an appropriate institutional framework for research and fundraising.

We find that most funders who contribute to science and research also contribute to other philanthropic causes including art, human services, other education (aside from higher education), and religion. To capture the diversity of focus across funding organizations, we identified 7,124,144 grants throughout all areas of philanthropy by funders who gave at least one grant to research institutions (science donors). We then identified for each funding organization the area to which it donates the largest amount of funds. Only 16% of science donors had an exclusive science focus and an additional 28% of science donors gave more to science than to any other area (Fig. S5b). Funders with a primary focus on science together account for 93% of all scientific philanthropy, suggesting that the bulk of science funding comes from organizations who have chosen it as their primary area of philanthropy. At the same time, we find several prominent funders who give to science, yet their primary focus area is elsewhere (Fig. S5). For example, the Annenberg Foundation primarily funds art organizations, and the Sherwood Foundation predominately funds primary and secondary education organizations, but both foundations also support scientific research related to art and primary education, respectively.

When we compare philanthropic grants to federal funding, we find that that in terms of the amount of funding, philanthropic support for institutions involved in research rivals the funding offered by the top national science funders in the US, with the combined total exceeding the yearly amount awarded by NSF and being comparable to the amount distributed yearly by the NIH (Fig. 2a). Indeed, we find evidence of funds explicitly designated for research or health amounting to nearly $4B/year in recent years (Fig. S3). Note that the increasing trend shown in Fig. 2a exaggerates the true rate of increase in philanthropic support, as the dataset has increasing coverage for the more recent years as more nonprofits filed online, and since as we discuss in SI Sec. II, only 30% to 40% of all funds donated to universities are earmarked specifically for research purposes. Yet, when we limit the data to organizations whose returns are included every year between 2013-2018, we continue to observe a 38% increase in philanthropic funding, indicating that foundations did considerably increase their support for science-related institutions in this period.

**THE DISTRIBUTION OF SCIENTIFIC GRANTS**

The non-profit ecosystem varies both in focus and might. We find that across the scientific philanthropic ecosystem, the total dollar amount of grants follows a heavy-tailed distribution (Fig. 2b), indicating that while most organizations distribute relatively small amounts, a few organizations devoted exceptional funds to research. For example, the Gates Foundation distributed over $6.5B in the past 10 years. Overall, the top 200 funders, corresponding to 0.3% of all grantmaking organizations, account for 66% of the total funds given to science. While major foundations dwarf smaller funders in terms of grant numbers and amounts, the long tail of the many smaller funders represents a considerable cumulative impact. For example, more than 7,000 funders have each donated at least $1M over the decade to scientific research institutions, levels of support that could be substantial for many programs.

Overall, we find that 13,000 organizations involved in science have received philanthropic support, comparable to the number supported by the NIH (15,000 organizations) or NSF (10,000 organizations)[25]. While federal support is highly concentrated on a few institutions, with the top 200 organizations receiving 80% of NIH and NSF grants, philanthropic support is spread more evenly, with the top 200 recipients attracting only around a third of the total philanthropic grants (Fig. S6e). Nonetheless, the distribution of the dollar amount received from philanthropic grants is largely indistinguishable from the distribution characterizing government grants (Fig. 2b) suggesting that while philanthropic grants might support more institutions, the bulk of the funds still go to a limited number of major recipients.

**PHILANTHROPY IS LOCAL**

A stated goal of federal funding for research is to support projects based on intellectual merit and broader impacts. While funding patterns often fall short of this goal, being affected by gender, racial, and other biases[26], federal funding aims to defy geographic boundaries. In contrast, as we show next, philanthropic funding is strongly affected by geography. We mapped each non-profit to its state of incorporation and identified grants distributed within the same state. If grants were distributed uniformly across the US, as expected in a merit-based system (preserving the number of grant recipients in each state), about 5% of grants are expected to be awarded in the home state of the donor (Fig. S9a). We find, however, that approximately 35% of grants go to the donor's state, a 7-fold increase over the random baseline. With 49% of funds remaining within the same state (after removing single-support foundations and donor-advised funds, see SI Sec. IV), compared to 4.5% expected by a national random model, we find the level of funding to be even more localized (Fig. 3a). The difference between the fraction of funds (49%) and the fraction of grants (35%) going in-state suggests that funders not only tend to give more grants locally, but also give larger grants to local recipients. Indeed, we find that for any particular funder their largest grant recipient is in the same state over 50% of the time, a proportion that decreases for smaller grants (Fig. 3d). While some foundations are explicit about their focus on local communities, most foundations lack such a mandate, hence, the local focus may be an unintended consequence of their limited ability to engage widely with the scientific community and the scientific community's failure to reach out to them.

Locality is reduced for funders who give more grants: while for organizations with a single beneficiary over 60% of the recipients are local, for organizations with more than 1000 recipients that fraction drops to 10% (Fig. 3e). Interestingly, funders who gave more money did not necessarily give less locally (Fig. S9e-f), suggesting that even those that give significant funds to research may focus their philanthropy on a few local recipients, while disregarding more distant institutions. Furthermore, while general, education, and scholarship grants all tend to be more local, we find that grants for research are still locally focused with 35% of dollars and 24% of grants remaining in-state (Fig. S3).

Unsurprisingly, large foundations with an explicit local focus and mandate tend to be even more locally focused, with the Lilly Endowment in Indiana distributing over 60% of its funds for science and research in the state of Indiana, the Sorenson Legacy Foundation in Utah giving 83% of its funds in-state and the Dennis Washington Foundation in Montana giving 99% of its funds in-state (see Fig. S8). Yet, this pattern is not limited to organizations with a local mandate. To systematically explore locality for funders explicitly focused on research, we identified 27 large foundations with a declared mission towards advancing scientific research (see SI section VIII). Often relying on formal calls that defy geographic boundaries, these foundations gave 35,389 grants worth $15.7B over the past decade. Despite their global infrastructure and mission, we find that these major science funders are still locally focused, giving, on average, 30% of their funds to organizations in the same state. For example, the Gates Foundation gives ten times more funds to science institutions within Washington State than expected based on a random geographic impact model (Fig. 3b-c). Similarly, the Ford and Rockefeller Foundations distributed three times more funds than expected in their home state of New York (Fig. S8). One of the least local funders is the Pennsylvania-based Templeton Foundation, which gives only 1.6 times more to recipients in its own state. In other words, even the largest foundations, who have the

infrastructure to seek out national and international applicants, tend to focus locally, either driven by a stated desire to impact their local communities or by unintended network effects, reflecting closer professional and social ties with local researchers and institutions.

**PHILANTHROPIC FUNDING IS STABLE**

Another dimension of grant giving relationships is donor retention[27,28], reflecting the likelihood of continued support given an already established funding relationship. We find that 69% of grant relationships repeat one year later (Fig. 4a) and 60% repeat two years later, compared to 8% repetition predicted in the random funding network. This high level of donor retention in the foundation space stands in strong contrast to online giving platforms where only 26% of grants repeated one year later[27]. Furthermore, repeated giving becomes increasingly entrenched over time, as donors who gave two years consecutively have an over 80% chance of giving the next year and for the 27,390 funding relationships that have been ongoing for 7 years there is a nearly 90% likelihood to continue in the next year (Fig. 4b).

We find that stable grants (ongoing between 2013-2019) are more likely to be given by organizations that offer fewer grants: over half of grantors who give to a single science recipient, support the same recipient every year. The fraction of stable recipients drops to 20% for those giving to a few dozen recipients (Fig. S14b). Similarly, stable grant relationships are more likely to occur when the donor supports the recipient at higher funding levels (Fig. 4c) and are more likely to occur with local relationships (Fig. S14g). Traditional science funders and funds specifically for research also exhibit stability, with 64% of grants repeating in the next year, and their long-term funding relationships also have a 90% likelihood to continue in future years (Fig. S16-S17).

10infrastructure to seek out national and international applicants, tend to focus locally, either driven by a stated desire to impact their local communities or by unintended network effects, reflecting closer professional and social ties with local researchers and institutions.

**PHILANTHROPIC FUNDING IS STABLE**

Another dimension of grant giving relationships is donor retention[27,28], reflecting the likelihood of continued support given an already established funding relationship. We find that 69% of grant relationships repeat one year later (Fig. 4a) and 60% repeat two years later, compared to 8% repetition predicted in the random funding network. This high level of donor retention in the foundation space stands in strong contrast to online giving platforms where only 26% of grants repeated one year later[27]. Furthermore, repeated giving becomes increasingly entrenched over time, as donors who gave two years consecutively have an over 80% chance of giving the next year and for the 27,390 funding relationships that have been ongoing for 7 years there is a nearly 90% likelihood to continue in the next year (Fig. 4b).

We find that stable grants (ongoing between 2013-2019) are more likely to be given by organizations that offer fewer grants: over half of grantors who give to a single science recipient, support the same recipient every year. The fraction of stable recipients drops to 20% for those giving to a few dozen recipients (Fig. S14b). Similarly, stable grant relationships are more likely to occur when the donor supports the recipient at higher funding levels (Fig. 4c) and are more likely to occur with local relationships (Fig. S14g). Traditional science funders and funds specifically for research also exhibit stability, with 64% of grants repeating in the next year, and their long-term funding relationships also have a 90% likelihood to continue in future years (Fig. S16-S17).

In terms of grant amounts, we find that funders who gave repetitively tend to give more money in the first and subsequent years of their support compared to donors who did not give repeatedly (Fig. 4d). Furthermore, the more years a relationship lasts, the larger the amount (Fig. 4e). At the same time, the typical grant relationship does not tend to involve an increasing donation amount and the median change in funding level after one, two, or even seven years of funding is near zero, indicating that donors give the same amount seven years later as they did in year one (Fig. 4f). This suggests that the initial amount a donor gives to a recipient sets the value of their overall relationship with that recipient and once the level of funding is established, donors rarely change their level of support.

**CLUSTERING IN PHILANTHROPIC SUPPORT**

The common statement used in philanthropy that "if you've met one funder, you've met one funder,"[29] implying that each philanthropic organization has its own unique and distinct priorities, hence understanding one funder's approach offers little information on the motivation of other funders. In contrast with this widely shared belief, we find evidence of strong clustering of funders in terms of their recipients, reflecting common focus and decision making (Fig. 5a). For example, Harvard received funds from 372 distinct foundations in 2019 and MIT received from 284, and 113 of these gave to both institutions, indicating that 40% of MIT donors and 30% of Harvard donors are in common. This strong overlap is not limited to prestigious universities but is a common feature of geographically proximal universities. The University of Nebraska Foundation and Creighton University (in Nebraska) shared 26 foundation donors in 2019, representing 35% of Creighton's donors and 17% of the University of Nebraska's donors. Among the overlapping donors we find several Nebraska-based foundations, such as the Robert Daugherty Foundation which gave $5.4M to the University of Nebraska and $1M to Creighton, and the Lozier Foundation which gave $594k to University of Nebraska and $295k to Creighton.

We quantify the donor overlap across the full philanthropic landscape via the bipartite clustering coefficient (see Methods),[30] allowing us to examine across all pairs of funders those who shared one recipient and the rate at which they shared other recipients. We find that the clustering coefficient for the bipartite donation network is 0.0448, which is 135 times the random baseline, meaning that two funders who shared one recipient are 135 times more likely to share another recipient than two randomly selected funders (see SI Sec. X). This strong funder overlap indicates that funders with similar funding priorities are driven to the same group of recipients.

**PREDICTABILITY IN PHILANTHROPY**

Finally, we ask whether the clustered nature of the grant network identified above carries inherent predictive power, which could be used to identify potential future funder-recipient pairs. Such predictions could assist those who seek funding to identify new funders, and help funders to diversify their portfolios by identifying organizations that focus on issues matching their mission. We focus on funders for the 3,279 science recipients who received funds from at least five distinct funders in 2018 and remained active in 2019, as well as 17,154 funders who were active in both 2018 and 2019. We used the bipartite Adamic-Adar Index (AA)[31,32] to predict the donors that are likely to donate to a specific recipient. This link prediction measure[32] suggests that if two recipients share some donors, then they are likely to share other donors, and that more unique shared recipients or donors convey more information. In other words, knowing that two universities are funded by a major funder like the Gates Foundation has less predictive value than knowing that they share other smaller funders. Explicitly, the Adamic-Adar index between a donor *s* and recipient *t* is given by,

$$AA_{s,t} = \sum_{\substack{paths\ of\ length\ 3 \\ from\ s\ to\ t}} \frac{1}{\log(|k_{i1}|+|k_{i2}|)}. \tag{1}$$

Donor-recipient (s,t) pairs with higher $AA_{s,t}$ scores are more likely to develop a funding relationship than pairs with lower scores (Fig. 5a). We convert the 2018 $AA_{s,t}$ scores to probabilities and test the model's predictive power by examining whether the predicted grant between a funder and a recipient was awarded in 2019. We find that the predictions obtained from the AA index from 2018 have strong predictive value for 2019, resulting in a remarkably high area under the receiver-operator curve (AUROC) of 0.87 (see SI Sec. X, Fig. 5b). An AUROC score of 0.5 indicates lack of predictive power and a score of one represents perfect predictions. The predictions, as measured by AUROC, remain equally good when we examine funding relationships above a threshold dollar amount ranging from $1-$10k, resulting in an AUROC between 0.87-0.90. For most research universities the leading prediction tends to be one of a few major funders like the Gates Foundation, Hewlett Foundation, Mellon Foundation, and Charles Koch Foundation, or corporate foundations like those of KPMG, Ernst & Young, or Shell Oil. Given that these foundations fund many universities, for any particular university there is a high likelihood of a grant. In contrast, for community colleges or smaller institutions that have limited access to national funders, the top predictions are often local funders, such as for the Anoka-Ramsey Community College in Minnesota, whose top predicted funder is the Minnesota-based Kopp Family Foundation, which indeed gave over $20k in 2019.

Next, we inspected cases where the AA index suggests high likelihood of a grant and yet no such grant exists in 2019, finding that these predictions tend to correspond to reasonable recommendations when we consider metadata and a timeframe of multiple years. For example, if we consider the top 100 donors predicted to donate at least $10k to Harvard in 2019, yet who did not, we find that 76 of them supported Harvard in a year other than 2019. A similar analysis for Creighton University reveals that 22 of the top 100 most-likely predicted donors gave in a year other than 2019. We also find that 18 of the top 100 predicted donors to Creighton who did not give were from the state of Nebraska, suggesting that while these donors may not have supported Creighton, it would be reasonable for them to do so given the previously discussed strong geographical patterns of funding. In Fig. 5d we show examples of two pairs of donors in Nebraska and Utah, along with the recipients ranked highest in their list of most likely funders. We see that several of these recipients previously received from the foundations and that many of the others are also local to the donor organizations. Furthermore, we see that the two Nebraska funders and the two Utah funders share many recipients who ranked them highly while none of the recipients ranked one of the other state's funders highly.

The observed high predictability[33] of future donors suggests that even though each funder has its own unique motivation, focus, governing structure and decision processes, there are common patterns in the funding outcomes. These common patterns can be exploited by organizations who seek funding for research, as well as by donors who aim for a better allocation of their funds, resulting in improved donor-recipient matches.

**DISCUSSION**

Our finding that philanthropic support for science is local aligns with other studies documenting the role of physical distance between funders and recipients in philanthropy[34-36]. Yet with 90% of published research papers written collaboratively, and 60% of publications listing authors from multiple institutions and multiple countries[37], modern science is an increasingly global pursuit that requires access not only to local, but to national and international talent and resources as well. The strong local focus of philanthropy documented above contrasts with these trends. At the same time, locality can increase equity by aiding organizations who lack the resources to fundraise nationally but have an established role and relationships within their local communities. But locality's effects also go in the other direction—while large philanthropic organizations acquired their funds through national and international investments, most of them are located in already affluent regions, hence their strong local focus can entrench existing geographic inequalities. If philanthropists aim to build the scientific capacity of their local region, then locality of funding may fit their mission, yet if the desire is to advance science itself, a local focus may be ineffective.

The implications of the documented stability in funding patterns are equally multifaceted. Indeed, scientific research can greatly benefit from stable funding, offering researchers the opportunity to take risks and focus on difficult problems that require long-term investments. At the same time stability may also represent inertia, rather than an intentional allocation of funds to further specific scientific or funding agendas. The inherent predictability of philanthropic relationships further raises questions of whether funders are seeking out distinct funding priorities or band together in supporting the same institutions due to network and/or prestige effects.[38] Given the exceptional amount of philanthropic wealth going towards science, it is worthwhile for philanthropic stakeholders to interface with the scientific community to achieve a better matching between donor intent and the supported research[3].

For individual researchers, the steady increase in philanthropic giving (Fig. 2a) offers increasing opportunities to seek funding beyond the federal funding system[39,40]. Yet, given their familiarity with the federal funding system, researchers tend to limit their fundraising efforts to those large private organizations that have a global presence and regular calls for proposals, operating similarly to federal funders. Our findings suggest, however, that there is exceptional value in engaging with local philanthropic communities, given the strong locality of funding patterns. Such local engagement could enable scientists to directly solicit support from philanthropists, rather than receiving indirect support from general grants that go to the institution's endowment. Local philanthropy is based more on relationships and outreach, rather than extensive proposals, and can offer more flexibility as philanthropists are not limited to supporting specific programs. Furthermore, the stable funding offered by philanthropists can advance projects with longer time horizons, not yet ripe for national or federal funders. Likewise, the stability of philanthropic funding could serve as an incentive for researchers to increase their scientific outreach efforts as philanthropic supporters of universities have been shown to donate more if they are able to more specifically direct their gifts[41]. Finally, the ability to predict funding relationships enables researchers to better identify and target philanthropists that are more likely to be interested in supporting their institutions, saving time and allowing them to focus their efforts.

Despite the exceptional amount of research and policy focus on national funding, there is limited quantitative understanding of philanthropic giving. Also, most of the existing knowledge relies on interviews and hand-curated datasets,[12-14] with advanced computational methods only beginning to enter the field[19]. Here we focused on funding information that can be extracted from US tax forms, offering a foundation for unbiased big-data-driven research to understand philanthropic giving and potentially improve access to philanthropic funds. While the richness of the dataset offered multiple insights, its limitations offer a roadmap for future data collection. First, the tax documents analyzed here are limited to the US, though other countries have also seen similarly increasing trends in philanthropic giving to science[42,43]. Second, these tax documents are limited to philanthropic giving by the approximately 80% of foundations that filed electronically (see SI Sec. I) and do not include giving from foundations who filed on paper or individual contributors. While the existing data somewhat underestimates private support for science, the increasing trends and requirements towards online filing will eventually alleviate this limitation. Furthermore, depending on the versions of the tax form, the employer identification number (EIN), a unique identifier, is not always available. We therefore relied on machine learning (SI. Sec. I) to identify recipients, with potential mismatches for a few organizations due to inconsistencies in the tax forms[44]. In addition, we were only able to disambiguate grants to other non-profits or entities for whom we have an EIN, hence we did not examine grants to non-US organizations and individuals. Finally, not all grants are equally impactful just as not all science is equally significant and some of the recorded grants, while they do support research institutions, only indirectly contribute to scientific research, facilitating instead infrastructure enhancements, undergraduate education, and administrative or programmatic tasks. At the same time, prior work has shown that general funding support does translate into research activity, as measured by publications and patents produced by a university[45]. Despite these and other limitations explored in depth in SI Sec. I, the tax data analyzed here offers the most comprehensive imprint of scientific philanthropy available over the past decade.

To aid the further use of this data for research, we are sharing the cleaned and organized data we extracted from the 990 forms (SI Sec. IX). The resulting dataset, amenable for data mining and for other research purposes, complements ongoing efforts by Candid, an organization that provides subscription-based searchable information for funders and scientific organizations. Further efforts are needed to expand this work to track philanthropic funders internationally who may fall under diverse tax laws with different types of reporting. Extensions of this work could help us better understand the nature of the science being funded, linking grants to individual scientists, publications, and patents, allowing researchers to explore the repercussions of locality and stability on scientific productivity and impact, as well as to develop quantitative measures to capture the efficiency of philanthropic and government funding. Such future work should also focus on policy implications, enhancing the relationship between stakeholders, from researchers to policy makers and funders. Through the application of novel tools rooted in machine learning, network science, and science of science[46,47], access to systematic philanthropic funding data could improve funding allocation, help organizations better provide for those they support, boost access to philanthropic resources, and enable policymakers to increase the impact of philanthropic funding.

**MATERIALS and METHODS**

*Data Collection and filtering.* Data was collected from AWS Open 990 filings at https://registry.opendata.aws/irs990/, note that since the end of 2021, the IRS hosts 990 filings directly on its website at: https://www.irs.gov/charities-non-profits/form-990-series-downloads. We identified all grants listed on donors' tax forms and for cases when only the recipient name and address were given, we applied a string-matching algorithm to determine a unique identifier for the recipient. We then filtered the set of grants down to those that went to organizations involved in science and research including universities and research institutions (see SI Sec. I). The final network in our study consisted of 69,675 donors and recipients and 926,124 grants for reporting years from 2010-2019.

*Filtering Special Cases.* Certain donors give large donations that can further bias the appearance of locality. While, across the entire dataset 67% of grant dollars were local, this includes many instances of a university having a separate foundation to receive grants and then making a large grant to the university annually. Therefore, for determining the fraction of grant dollars given locally, we filtered such foundations and other edge cases such as the NCAA, and donor-advised funds resulting in the 49% of dollars donated locally mentioned in the main text. See SI Sec. IV for more on this.

*Bipartite Clustering Coefficient.* To measure the bipartite clustering coefficient we used the definition from Robins and Alexander,[30]

$$C(x) = \frac{4 * C_4}{L_3}, \qquad (2)$$

where *C* is the clustering coefficient, $C_4$ is the number of cycles of length four and $L_3$ is the number of paths of length three.

*Data Availability*. The final resulting network of science grants is available at https://github.com/Barabasi-Lab/mapping-philanthropy/. See SI Sec. IX.

**ACKNOWLEDGEMENTS**

We thank Danielle Lamay for assisting with the data cleaning. We thank Dimensions.ai for providing access to their database and acknowledge useful discussions with George Overholser, Bingsheng Chen, Larry Mcgill, Jacob Harold, Anna Koob, and other individuals at Candid.

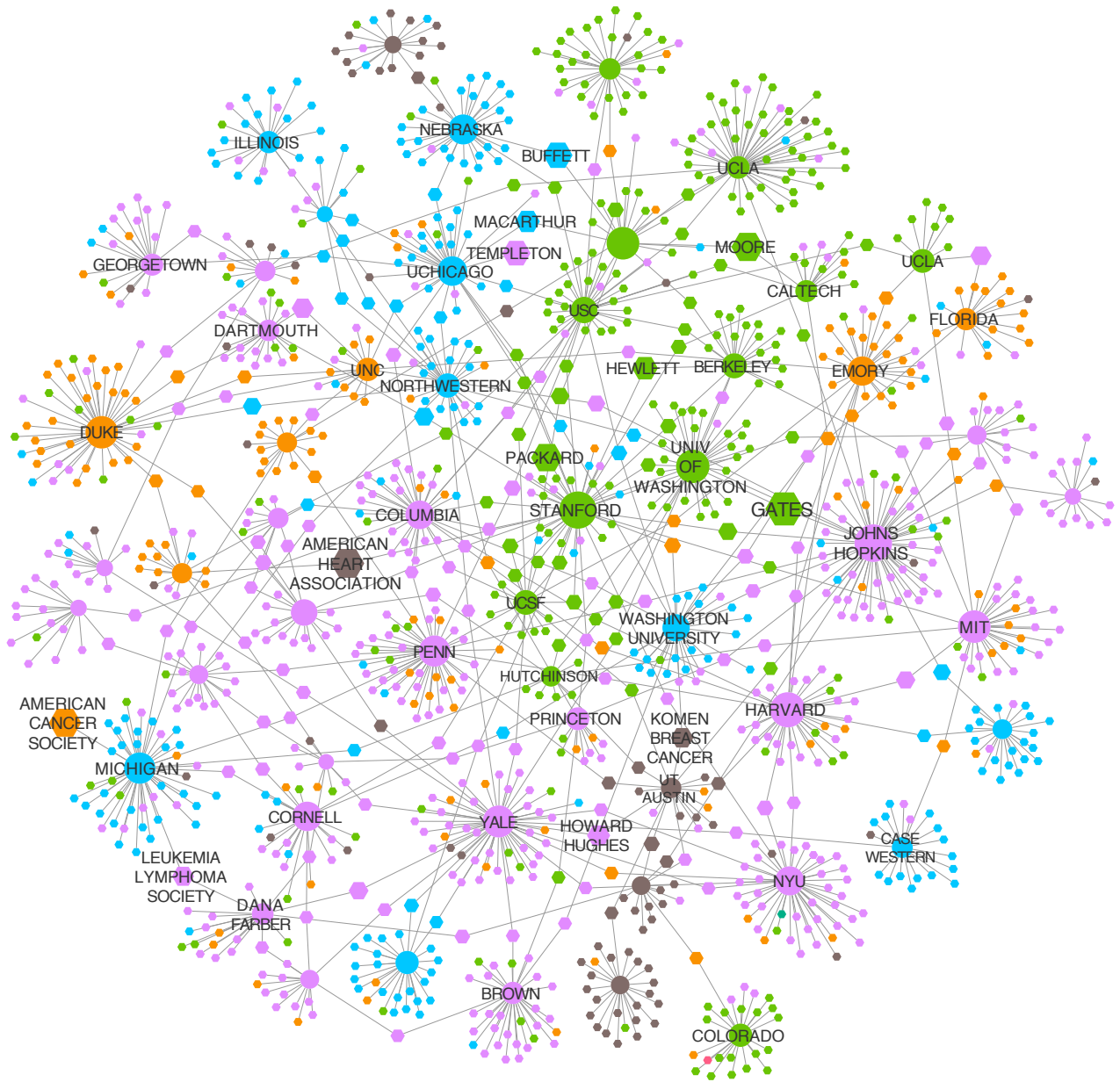

Fig. 1. **The Philanthropic Ecosystem of Science.** The network of funders and their top recipients. For each funder we maintained their top two recipients and then filtered to only include relationships worth over $1M over the period examined. We also removed donor advised funds and single-support foundations. The resulting network shows 55 recipients (circles) and 1254 donors (octogons) with 1422 grant relationships between them. Nodes are colored by region with purple being Northeast, blue being Midwest, green being west, brown

being southwest, and orange being the south. We see that most donors have their top recipient/s in the same region, though those with multiple $1M+ recipients at times have a top recipient in a different region.

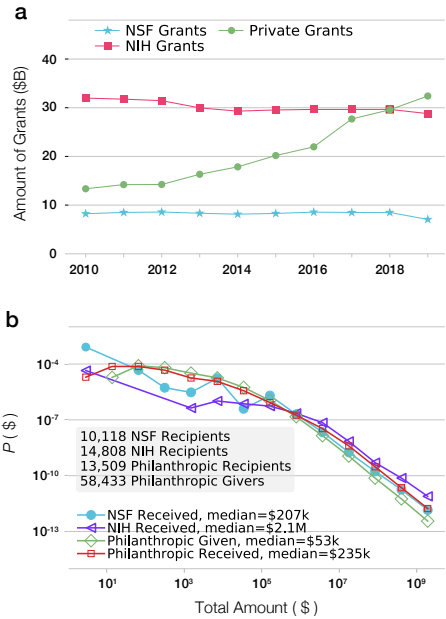

Fig. 2. **Philanthropic vs. Federal Support. (a)** The amount of grants provided to institutions performing research by private nonprofit organizations has grown considerably over the past decade, surpassing the amount of grants given by the NSF and NIH [14]. **(b)** The distribution of the total amount of science-related grants given or received by philanthropic organizations, compared to grants distributed by NSF and NIH. Note that while the NSF, NIH, and philanthropy all support similar numbers of recipients, there are far more philanthropic donors than recipients.

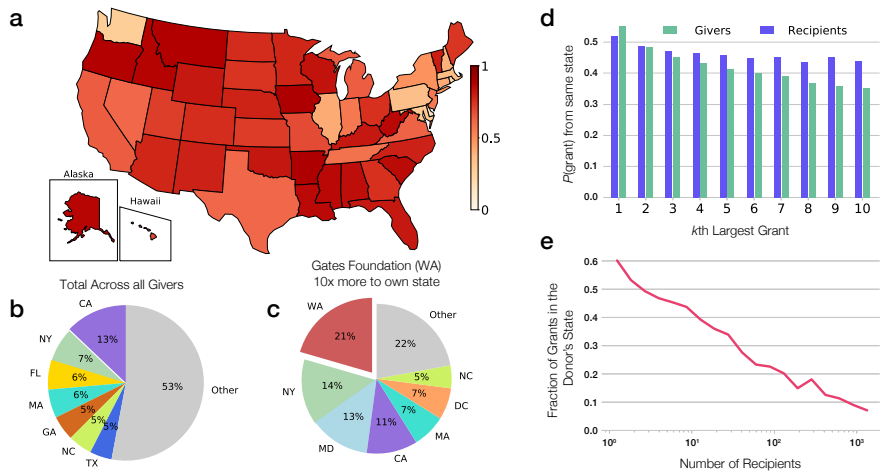

Fig. 3. **Locality in Philanthropy**. **(a)** The fraction of dollars given within the state for donors in each state. **(b)** The total cumulative proportion of dollars going to particular states when all donors are considered. Particular grantors are seen to be focused towards their individual states including large funders such as **(c)** the Gates Foundation located in Washington. **(d)** For donors and recipients, we show the likelihood that their *k*th largest recipient was in the same state. For donors we see a decreasing trend, indicating that the largest recipient is more likely to be in the same state than recipients who received less funds. For recipients, while their largest donor is somewhat more likely to come from the same state, the decline for smaller donors is much slower. **(e)** The fraction of grants given within the donor's home state as a function of the number of recipients supported by the donor. Givers with fewer recipients tend to give more locally compared to those with more recipients.

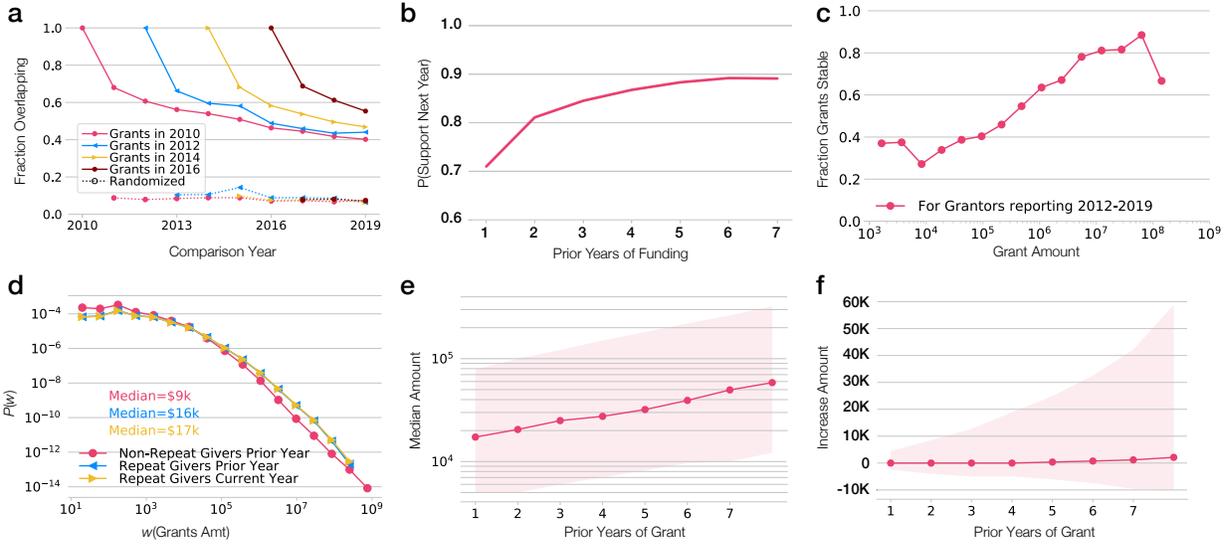

Fig. 4. **Stability in Philanthropy**. **(a)** For a particular year (2010, 2012, 2014 & 2016), we plot the fraction of grants that overlap in future years, a signature of continued support to the same recipient by a funding organization. In the randomized versions of the philanthropic network the overlap is under 10%. **(b)** The likelihood that a funder will continue supporting a recipient as a function of the number of years of prior support. Grant relationships tend to become increasingly entrenched over time as longer relationships are more likely to continue than shorter ones. **(c)** The fraction of stable grants (continuing for 7+ prior years) versus the grant amount in 2019. The increasing trend suggests that larger donors are more likely to have a stable relationship with their recipients. **(d)** The distribution of grant amounts for grants that do not repeat, the prior year of a repeating grant, and the current year of a repeating grant. The fraction of dollars given within the state for donors in each state. **(e)** The median amount of a grant this year as a function of the number of prior years the grant relationship has existed. **(f)** The median change in grant amount as a function of the number of prior years of the grant relationship.

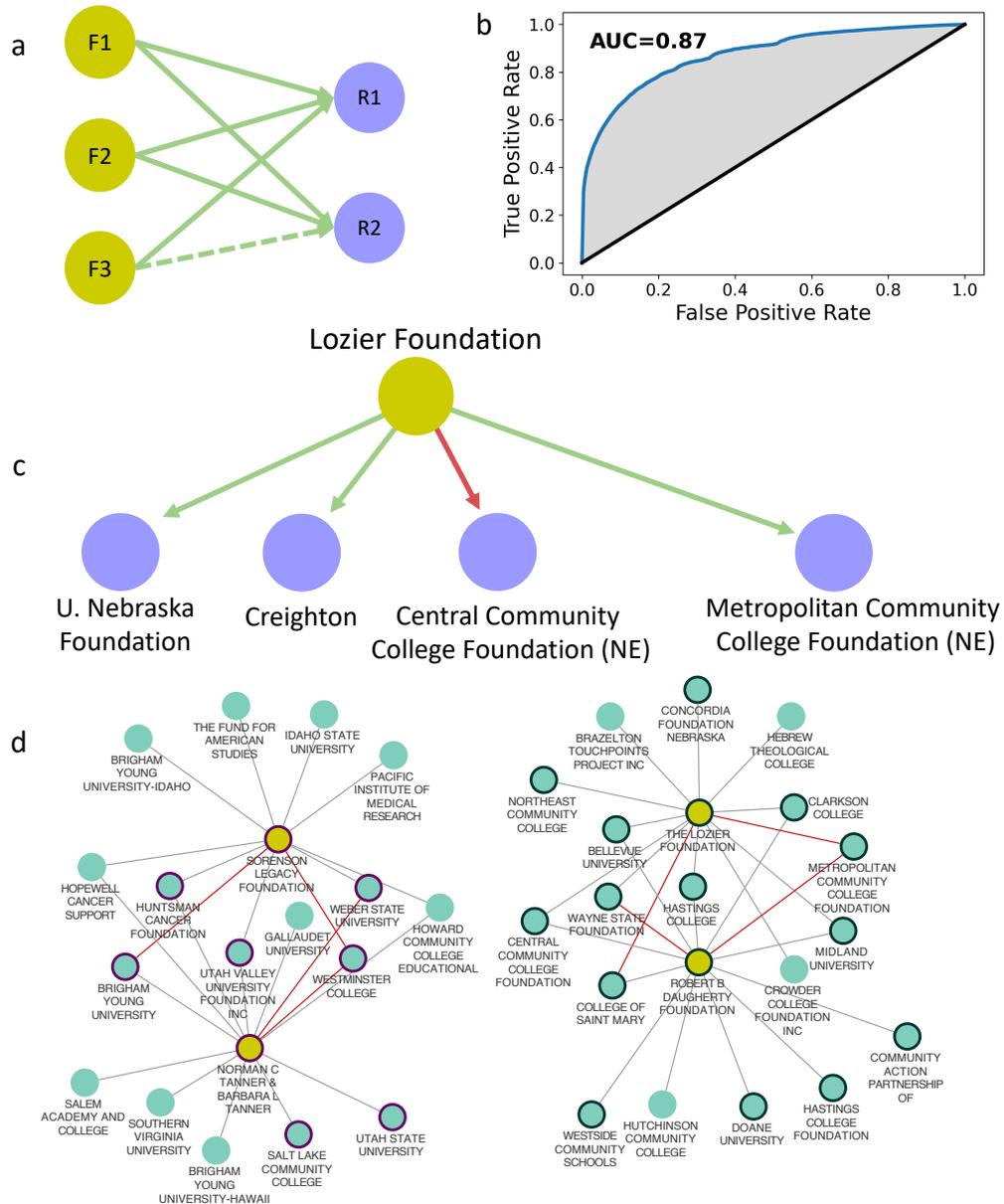

Fig. 5. **Predictability of Philanthropic Grants**. We use the bipartite Adamic-Adar index to measure the similarity and predictability of grant relationships based on the network of all funder-recipient relationships. **(a)** A demonstration of the link prediction approach. If funder F1 and F2 both supported recipients R1 and R2, then funder F3, who supports recipient R1, is more likely to support recipient R2 as well. **(b)** The receiver-operator curve (ROC) for predictions using the AA index from 2018 on the network of grants over $10,000 to predict grants over $10,000 in 2019. The area under the curve (AUC) is 0.87. **(c)** The top predicted recipients (from left to right) of the Nebraska-based Lozier Foundation and whether Lozier supported them (link is green if it did, red if it did not). The two highest predicted recipients were the University of Nebraska Foundation and Creighton who did receive support from the foundation, as did the fourth highest predicted recipient, the Metropolitan Community College Foundation. The third highest predicted recipient, the Central Community College Foundation did not receive from Lozier in 2019. **(d)** Examples of resulting predictions. For four funders (in gold) we show the recipients for whom they were highest ranked (see SI note). On the left we show the resulting prediction network for the Sorenson Legacy Foundation and Tanner Foundation, both located in Utah. We highlight recipients in Utah with a purple border. On the right we show the recipients who ranked the Lozier and Daugherty foundations highest and highlighted with a black border those in Nebraska. We see that the network structure alone identified predictions consistent with the locality of grants.

39    McNutt, M.  Vol. 344   9-9 (American Association for the Advancement of Science, 2014).
40    Ledford, H. Sponsor my science: philanthropists will sometimes give large sums of money to support science--but researchers have to learn how to sell themselves first. *Nature* **481**, 254-256 (2012).
41    Eckel, C. C., Herberich, D. H. & Meer, J. A field experiment on directed giving at a public university. *Journal of behavioral and experimental economics* **66**, 66-71 (2017).
42    Gouwenberg, B. *et al.* Foundations supporting research and innovation in Europe: results and lessons from the Eufori study. *The Foundation Review* **8**, 11 (2016).
43    Kundu, O. & Matthews, N. E. The role of charitable funding in university research. *Science and Public Policy* **46**, 611-619 (2019).
44    Gordon, T., Khumawala, S. B., Kraut, M. A. & Meade, J. A. The quality and reliability of Form 990 data: Are users being misled. *Academy of Accounting and Financial Studies Journal* **11**, 27 (2007).
45    Tabakovic, H. & Wollmann, T. G. The impact of money on science: Evidence from unexpected NCAA football outcomes. *Journal of Public Economics* **178**, 104066 (2019).
46    Fortunato, S. *et al.* Science of science. *Science* **359**, eaao0185 (2018).
47    Wang, D. & Barabási, A.-L. *The science of science*.  (Cambridge University Press, 2021).
30